\documentclass[aps,pra,twocolumn,groupedaddress,10pt]{revtex4-1}

\usepackage{
amsmath,
amsfonts,
amssymb,
dsfont
}
\newcommand{\phivev}{\langle\phi\rangle}
\newcommand{\sigmavev}{\langle\sigma\rangle}

\begin{document}

\title{Vanishing trace anomaly in flat spacetime}

\author{Zygmunt Lalak}
\email[]{Zygmunt.Lalak@fuw.edu.pl}

\author{Pawe{\l} Olszewski}
\email[]{Pawel.Olszewski@fuw.edu.pl}

\affiliation{Institute of Theoretical Physics, Faculty of Physics, University of Warsaw ul. Pasteura 5, 02-093 Warsaw, Poland}

\date{\today}

\begin{abstract}
Quantum scale invariant regularization is a variant of dimensional regularization where the renormalization scale is treated as a dynamical field. But, rather than be regarded as a novel regularization method on par with dimensional regularization, momentum cutoff, Pauli-Villars etc., it should be understood as a way to define a subset in the infinite space of nonrenormalizable models of certain type. The subset realizes the demand that renormalization scale, along with any other dimensionful parameters, should be interpreted as a dynamical field's homogeneous background. This restriction is most straightforwardly implemented using dimensional regularization but it can hypothetically be imposed with any regularization method. Theories that satisfy it offer a new perspective on the radiative violation of global scale symmetry associated with RG functions. As a result of the quantum scale invariant regularization being implemented, the scale symmetry is preserved at the quantum level despite the RG functions being nonzero, as can be inspected at the level of composite quantum operators that govern dilatation of Green functions. We analyze these statements in explicit detail using a specific but easily generalized toy model with scalar fields.  
\end{abstract}


\maketitle

\section{Introduction}
Our goal in this article is to explain in detail and justify the following claim. A model with a massive scalar field with a nonzero vacuum expectation value can be embedded in a nonrenormalizable theory with one additional scalar singlet field in such a way that the resulting theory is, first, classically scale symmetric and, second, free of the so called \textit{trace anomaly}. The latter characteristic must be accompanied by a disambiguation. The \text{trace} refers to the trace of renormalized composite quantum operator that is the quantum counterpart of energy-momentum tensor known from classical physics. We summarize some relevant elementary properties of the energy-momentum tensor in the Appendix \ref{appendix}, the Reader may wish to consult it before engaging with the main body of the article. In a perturbative scheme one can talk about presence of the classical scale symmetry that is violated by small quantum corrections. This description is paralleled by calculation of the said \textit{trace} which is zero at zeroth order but receives loop corrections. The latter can be classified into two subsets. First, Weyl (or conformal) anomalies which happen to vanish in flat space-time. They are the genuine anomalies in the sense that they originate from a nontrivial jacobian produced by treating a local scale transformation as a change of integration variables in a path integral (see e.g. \cite{Deser:1993yx, fujikawa2004path} and references therein). Second, there is the contribution independent of the metric that stems from the presence of a dimensionful renormalization scale in the formalism. This contribution is related to nontrivial RG functions. It is only this second contribution that we will refer to as the \textit{trace anomaly}.

The statement whether or not a symmetry of a given model is anomalous has to be independent of regularization method employed in dealing with loop divergences. For example, Lorentz symmetry or supersymmetry, when they are implemented as global symmetries of a quantum field theory, are generically non-anomalous regardless of what regularization is used in, say, intermediate steps of calculating renormalized correlation functions. On the other hand, using a regularization that explicitly preserves some or all of the hypothesized physical symmetries offers practical advantage even if a symmetry ends up being anomalous. It is not an accident that dimensional regularization has become the most popular method in gauge theories or that calculation of the $\pi^0 \to \gamma \gamma$ width and the associated chiral anomaly had remained clouded for some years.

Keeping the above in mind we wish to elucidate a certain \textit{manner} of regularization. The manner explicitly preserves scale invariance in a certain sense, and it has been proposed and used in the literature for some time.

Consider the Poincare algebra extended by dilatations $D$ (see Appendix \ref{appendix}).
In short, the main idea is to use dimensional regularization and to treat the renormalization scale $\mu$ as a function of a scalar field (or fields) that can be expanded around the field's background value, often but not necessarily taken to be the field's vacuum expectation value,
\begin{equation}
\mu \,\to\, \mu(\Phi) = \mu(\langle\Phi\rangle) + \text{interactions}
\end{equation} 
This way $\mu$, rather than being a dimensionful but inert parameter with respect to dilatations,
$
i[D,\mu] = 0
$,
 derives its dimensionality from a quantum field, so that
\begin{gather}
 i[D,\mu] = \frac{\partial\mu}{\partial \Phi} i [D,\Phi] = (1+ x^\nu\partial_\nu)\mu
\end{gather} 
The freedom of switching between renormalization schemes is still realized by arbitrariness of the $\mu(\Phi)$ function. But crucially, connection between choosing a renormalization scheme and introducing a dimensionful parameter, the so called renormalization scale, is severed.

Some of the earlier work concerned with conformal anomalies or lack thereof depending on the choice of renormalization scheme includes the articles \cite{Englert:1976ep,Fradkin:1978yw} (see also references therein) where the general coordinate transformation invariance along with Weyl invariance are considered. Later the interest in scale symmetry in curved background was revived by Shaposhnikov \textit{et. al.} in the line of research that utilizes the regularization scale $\mu$ as a function of dynamical fields in constructing cosmological models; see \cite{Shaposhnikov:2008xb, Shaposhnikov:2008xi} accompanied by \cite{Shaposhnikov:2009nk} (where the latter examines the interplay between choice of $\mu$ and nonrenormalizability), continued in \cite{GarciaBellido:2011de,Bezrukov:2012hx,Karananas:2016kyt} and also later in \cite{Ferreira:2016vsc,Ferreira:2016wem,Ferreira:2016kxi,Ferreira:2018itt,Ferreira:2018qss}. Those works are mainly focused on phenomenology of the early Universe but they rely on vanishing of the trace anomaly in curved background.
In the current article we consciously limit the scope of discussion to a theory in flat space-time.
A considerable part of the article discusses the same issues and conclusions as \cite{Tamarit:2013vda}. Here however we do not rely on referencing the quantum action principle for the sake of a more hands-on derivation and we aim to place the concept of 'scale invariant manner of regularization' in the broader context of quantum field theory in a pedagogical manner. We also hope to offer an useful point of entry to the formal construction that underlies the analysis in \cite{Ghilencea:2015mza, Ghilencea:2016ckm} (the latter coauthored by the current authors) that was concerned with explicit calculation of loop diagrams within the setup presented here.

In the first part of the article, Secs. \ref{spuriositySection}-\ref{roleofmuSection}, we define this regularization procedure and discuss its relation to a standard approach. We give an example of a model regularized in this scale invariant way and discuss how it relates to more general models renormalized in the standard way that employs the MS scheme and dimensional regularization. We use a minimal toy-model with scalar fields as an illustration albeit one which is immediately relevant for extending the standard model \cite{Tamarit:2013vda,Ghilencea:2016dsl}. In the second part, Secs. \ref{Quantum-scale-invariance-section}-\ref{special-conformal-section}, we discuss in detail, at the level of regularized composite operators, how models that comply with the scale invariant regularization do not have the dilatation symmetry violated by radiative corrections in flat space-time and are thus free of the trace anomaly. We derive this statement explicitly for the discussed toy model.

\section{Background fields and spuriosity}
\label{spuriositySection}
Recall the formal route from a classical action $S[\phi]$ to quantum effective action $\Gamma[\phi]$, the 1PI generating functional, \cite{Pokorski:1987ed},
\begin{gather}
Z[J] = -i \log \int \mathcal{D} \phi \, \exp \left( iS[\phi] + i J\cdot \phi \right) \;,
\\
\text{and} \quad
\Gamma[\phi] = Z[J[\phi]] - \phi \,J[\phi]\; ,\quad \text{where}
\\
J[\phi] \quad \text{ is the inverse of } \quad \phi[J] \equiv \frac{\delta Z[J]}{\delta J}
\end{gather}
and the straightforward extension of those definitions by a background field configuration $\phi_\text{bg}(x)$ introduced for a dynamical field $\phi$ in the following way
\begin{gather}
Z(\phi_\text{bg})[J] = -i \log \int \mathcal{D} \phi \, \exp \left( iS[\phi+\phi_\text{bg}] + i J\cdot \phi \right)\;, \\
\text{and} \;\;
\Gamma(\phi_\text{bg})[\phi] = Z(\phi_\text{bg})[J_{\phi_\text{bg}}[\phi]] - \phi \,J_{\phi_\text{bg}}[\phi]
\, ,\;\; \text{where}
\\
J_{\phi_\text{bg}}[\phi] \;\; \text{ is the inverse of } \;\; \phi_{\phi_\text{bg}}[J] \equiv \frac{\delta Z(\phi_\text{bg})[J]}{\delta J}
\end{gather}
From those definitions it directly follows that
\begin{equation}
\Gamma(\phi_\text{bg} + \Delta \phi_\text{bg})[\phi] = \Gamma(\phi_\text{bg})[\phi+ \Delta \phi_\text{bg}]
\label{bgshift}
\end{equation}
for any function $\Delta \phi_\text{bg}$.

Any effective action $\Gamma$ can be regarded as equipped with this additional dependence on background fields, regardless of whether one chooses to use it or not. Further, this dependence may be restricted to only homogeneous backgrounds.
\begin{equation}
\Gamma(\phi_\text{bg}) \to \Gamma(\phi_\text{bg})\Big|_{\displaystyle \phi_\text{bg} = \phivev = const}
\end{equation}
(We will use $\phivev$ as a single symbol to refer to a generic constant value with appropriate dimension, and not necessarily the vacuum expectation value of the field $\phi$. Later, $\phi_0$ shall denote a bare field rather than a constant. Similar remarks apply to $\sigma$.)

We will be assuming that the effective action may be written as a spacetime integral of an effective Lagrangian, as is the case when the expansion in powers of field derivatives is applied,
$\displaystyle
\Gamma[\Phi] = \int \!\mathrm{d}^4 x\, \mathcal{L}_\text{eff}(\phi,\partial \phi,\partial^2 \phi,\ldots)
$. 
$\mathcal{L}_\text{eff}$ is a sum of local interactions accompanied by number valued parameters, dimensionless coupling constants $\{\lambda\}$ and mass parameters $\{m,M\}$ with positive dimension.
How is one to distinguishes between a mass parameter $m$ and a constant background $\phivev$ given that they are both arguments in $\Gamma(m,\phivev)$?
We adopt \eqref{bgshift} as the defining property of a background field that provides the criterion. If
\begin{align}
\begin{split}
&\mathop{\forall}_{\Delta\phivev = const} \Gamma(m+\Delta\phivev,\phivev)[\phi]
\\
&= \Gamma(m,\phivev+\Delta \phivev)[\phi] = \Gamma(m,\phivev)[\phi+\Delta\phivev]
\label{spuriosity}
\end{split}
\end{align}
then we will refer to the parameter $m$ as \textit{spurious} with respect to the field $\phi$ and we will drop dependence on $m$ from the functional $\Gamma$ via
$\displaystyle
\Gamma(m,\phivev) \to \Gamma(0,\phivev) \equiv \Gamma(\phivev)
$,
where the rightmost functional is formally a new entity with one less parameter. We will refer to this mapping as absorption of the spurious $m$ by the field $\phi$.

The following observation is crucial for us. If a parameter is spurious at the level of a classical action, it will be spurious in the effective action as well
\begin{equation}
\mathop{\forall}_{\Delta\phivev = const} S(m+\Delta\phivev)[\phi] = S(m)[\phi+\Delta \phivev] \;\Rightarrow\; \text{Eq.}\eqref{spuriosity}
\label{implication}
\end{equation}
This implication holds for bare effective action and may in principle be spoiled by an unsuitable choice of renormalization scheme. Consider for example $\mathcal{L} = \frac{1}{2}(\partial \phi)^2 - V(\phi)$, where
\begin{equation}
\quad V(\phi) = \Omega + j\phi + m^2 \phi^2 + y\phi^3 + \lambda\phi^4
\label{sillyaction}
\end{equation}
If we demand that $\Omega = \lambda v^4$, $j = 4\lambda v^3$, $m^2 = 6\lambda v^2$ and $y = 4\lambda v$, then
\begin{equation}
V(\phi) = \lambda(v+\phi)^4
\label{sillyaction2}
\end{equation}
at which point the parameter $v$ becomes spurious with respect to $\phi$. In such case we unequivocally change the interpretation of $v$. It is absorbed by $\phi$ and replaced with $\phi$'s background value, $V=\lambda(\phivev+\phi)^4$, where $\phivev$ is regarded as charged under the same symmetries as $\phi$. The resulting action has no dimensionful parameters other than a renormalization scale.

From the point of view of the five-parameter Lagrangian in \eqref{sillyaction}, expressing $\Omega$, $j$, $m^2$ and $y$ as functions of a single dimensionful parameter $v$ and demanding it to be spurious in the resulting effective action constitutes a constraint on the choice of the model. By arranging for this constraint to be respected, we find ourselves considering a more symmetric physical setup. Thus, in typical terms, the potential in \eqref{sillyaction2}, despite the fact that it simply refers to a subset of potentials in \eqref{sillyaction}, describes a new, qualitatively distinct \textit{model}. Let us clarify and stress this point. A \textit{model} is first and foremost defined by the choice of quantum fields and their interactions encoded in the Lagrangian. If, in the practice of model building, one analyzes a range of numerical values for coupling constants or masses, each individual point is usually not referred to as a separate \textit{model}, although technically it should. This semantic distinction becomes more useful when, by changing values of parameters, one comes close to a subspace of the parameter space where some new symmetry is realized. (One can easily come up with relevant examples: a \textit{model} with an appropriate number of fields with different spins may become supersymmetric upon imposing a specific relation between its couplings, a \textit{model} with massive vector bosons is seen to realize the Higgs mechanism not only thanks to an appropriate particle spectrum but also due to respecting a set of constraints between couplings and masses; and so on.) Once we find ourselves within the subset of more symmetric models (or even just close to it), presence of the symmetry (perhaps approximate or anomalous) also becomes part of what defines the \textit{model} under consideration. Thus, the potential in \eqref{sillyaction2} describes a different \textit{model} than the one in \eqref{sillyaction} precisely because it realizes scale invariance at the tree level. Notice that this classification relies on our ability to spot the spuriosity of parameter $v$ in \eqref{sillyaction2}. Henceforth we will stop to italicize the word \textit{model} even though it will refer to a class of models defined broadly by a choice of fields, their interactions and (almost) realized symmetries and not necessarily a choice of specific numerical values for couplings etc.

In summary:

\paragraph{}
Every 1PI effective action functional depends, implicitly or explicitly, on background fields.

\paragraph{}
By a suitable choice of parameters in the Lagrangian some dimensionful parameters of the action may become \textit{spurious} i.e. absorbable in the backgrounds, effectively vanishing from the functional. We will refer to such models as obeying a spuriosity constraint, with implicit reference to the relevant parameters. 

For completeness we also add two remarks:

\paragraph{}
Given a stationary background $\phi_\text{bg}^\text{eom}$ such that
$
\displaystyle
0\!=\!\frac{\delta \Gamma(\phi_\text{bg}^\text{eom})[\phi]}{\delta\,\phi}\bigg|_{\phi=0} \!,
$
the restricted functional $\Gamma_\text{eff} \equiv \Gamma(\phi_\text{bg}^\text{eom})[\phi\!=\!0]$ coincides with the product of integrating the field $\phi$ out in the sense of $\int \mathcal{D}\psi \,e^{\Gamma_\text{eff}[\psi]} = \int\mathcal{D}\psi\mathcal{D}\phi\, e^{\Gamma[\phi,\psi]}$.

\paragraph{}
One may arbitrarily promote a parameter (irrespective of its dimensionality) to the background of a hypothetical field absent from the action and regard this parameter as belonging to a representation of global symmetries along with proper fields. Such parameter is called a spurion and the task of spotting all interesting symmetries that employ spurions is known as spurion analysis. We would like to stress the distinction between spurion analysis and the notion of restricting a dimensionful parameter to become spurious followed by absorption into a field, as introduced above. The latter is possible only when the relevant field is actually present in the dynamics.

\section{Absorbing the renormalization scale}
\label{absorbingSection}

Discussion in the previous section allows us to succinctly state the first main claim of this note: the renormalization scale can be a spurious parameter. Whether it is or not, depends on the choice of fields' spectrum, their interactions and the renormalization scheme. We conjecture that it does not depend on what method of regularization is used. Still, we will provide only an example that uses dimensional regularization. The reason is that we would like a method that explicitly introduces renormalization scale into classical action in order to use the implication in \eqref{implication}. Consider the following $d=4-2\epsilon$ dimensional model of two real scalar fields $\phi$ and $\sigma$ of dimension $1-\varepsilon$,
\begin{gather}
\mathcal{L} = Z_\phi\frac{1}{2}(\partial\phi)^2 + Z_\sigma \frac{1}{2}(\partial \sigma)^2 \,-\,
 V(\phi,\sigma)
\label{thelagrangian}
\end{gather}
\begin{gather}
\begin{split}
V(\phi,\sigma) \,=\, &\left[e^t(\sigmavev + \sigma)^\frac{1}{1-\varepsilon}\right]^{2\varepsilon}
\\
&\times \sum_{n=0}^\infty Z^{(n)} \lambda^{(n)} \,\phi^{2n}\,(\sigmavev + \sigma)^{4-2n}
\end{split}
\label{potential}
\end{gather}
 where each of the terms in the sum above with $n>2$ is interpreted as an infinite power series of nonrenormalizable interactions,
 \begin{widetext}
\begin{equation}
 \lambda^{(n)}\phi^{2n}(\sigmavev+\sigma)^{4-2n}  = \lambda^{(n)}\frac{\phi^{2n}}{\sigmavev^{2n-4}}
 \left[ 1 - (2n-4)\frac{\sigma}{\sigmavev} + \frac{(2n-4)(2n-3)}{2}\frac{\sigma^2}{\sigmavev^2} + \ldots\right]\;,\quad \left|\frac{\sigma}{\sigmavev}\right|<1
\end{equation}  
 \end{widetext}
 and so we restrict ourselves to a \textit{broken phase} of any symmetry realized on the field $\sigma$, i.e. $\sigmavev > 0$.
We will justify below, why all these terms have to be included.
The $Z$ factors contain perturbative counterterms, including the necessary poles in $\varepsilon \to 0$ and finite terms that define the renormalization scheme. 
Clearly, the $1-\varepsilon$ dimensional parameter $\sigmavev$ is spurious with respect to the field $\sigma$ at the tree level. Based on our earlier discussion, we claim that it will remain spurious in the effective action derived from this Lagrangian using for example the minimal subtraction scheme.

To better understand the Lagrangian in \eqref{thelagrangian} we observe that it is a special case of a general nonrenormalizable model of $\phi$ and $\sigma$ that can be written as
\begin{gather}
\widetilde{\mathcal{L}} \,=\, Z_\phi\frac{1}{2}(\partial\phi)^2 + Z_\sigma \frac{1}{2}(\partial \sigma)^2 \,-\,\widetilde{ V}(\phi,\sigma)
\label{lagrangian2}
\end{gather}
\begin{gather}
\begin{split}
\widetilde{ V}(\phi,\sigma) \,=\, \sum_{n=0}^\infty \sum_{m=0}^\infty
&\left(e^t \mu_0\right)^{\varepsilon(2n+m-2)}
\\
&\times
 \widetilde{Z}^{(n,m)} \widetilde{\lambda}^{(n,m)}\,\frac{\phi^{2n}\sigma^m}{M^{2n+m-4}}
 \label{thelagrangian2}
 \end{split}
\end{gather}
The renormalization scale $\mu\equiv e^t \mu_0$ is arbitrarily split into a fixed mass parameter $\mu_0$ and dimensionless RG factor $e^t$. After choosing $\mu_0=M$ the effective action computed from $\widetilde{\mathcal{L}}$ may be restricted by the requirement that $M$ is spurious with respect to the field $\sigma$. Deriving such restricted action directly from \eqref{thelagrangian} is simply a convenient technical shortcut, a trick to guarantee that the spuriosity constraint is respected.

Yet, upon closer inspection, one may object against identifying the models in \eqref{thelagrangian} as a subset of those in \eqref{thelagrangian2}. Consider the interaction terms in the former,
 \begin{widetext}
\begin{align}
\mathcal{L} \;\supset\,& 
\sum_{n=0}^\infty 
(e^t)^{2 \varepsilon} (\sigmavev^\frac{\varepsilon}{1-\varepsilon})^{2+(2n-4)} \;Z^{(n)} \lambda^{(n)} \;
 \frac{\phi^{2n}}{\sigmavev^\frac{2n-4}{1-\varepsilon}} 
\left(1+\sigmavev^\frac{\varepsilon}{1-\varepsilon} \frac{\sigma}{\sigmavev^\frac{1}{1-\varepsilon}} \right)^{4-2n+\frac{2\varepsilon}{1-\varepsilon}} \\
=& \sum_{n=0}^\infty
(e^t)^{2 \varepsilon} (\mu_0^\varepsilon)^{2n-2}\; Z^{(n)} \lambda^{(n)} \;
\frac{\phi^{2n}}{M^{2n-4}}
\underbrace{\left(1+\mu_0^\varepsilon \frac{\sigma}{M}\right)^{4-2n+\frac{2\varepsilon}{1-\varepsilon}}}_{\equiv \sum_{m=0}^\infty \left(\mu_0^\varepsilon\frac{\sigma}{M}\right)^m a_{n,m}}
\Bigg|_{\mu_0=M=\sigmavev^\frac{1}{1-\varepsilon}} \\
=& \sum_{n=0}^\infty \sum_{m=0}^\infty
(e^t)^{2 \varepsilon} (\mu_0^\varepsilon)^{2n+m-2} \; Z^{(n)}\lambda^{(n)} a_{n,m} \; \frac{\phi^{2n}\sigma^m}{M^{2n+m-4}} \Bigg|_{\mu_0=M=\sigmavev^\frac{1}{1-\varepsilon}}
\label{thelagrangian3}
\end{align}
 \end{widetext}
and compare them with the potential $\widetilde{V}$ in \eqref{thelagrangian2}. There are two notable observations.

\paragraph{}
Each $Z^{(n)}\lambda^{(n)} a_{n,m}$ factor represents a constrained choice of $\widetilde{Z}^{(n,m)} \widetilde{\lambda}^{(n,m)}$. Notably, aside from poles and strictly finite terms, it also contains terms with positive powers of $\varepsilon$ that we have hidden in the coefficients $a_{n,m}$,
\begin{align}
\label{anm}
&\left[a_{n,m}\right]\,=
\\
\nonumber
&\renewcommand{\arraystretch}{1.2}
\setlength\arraycolsep{7pt}
\left[
\begin{array}{llll}
 1 &
 4+2 \varepsilon +2 \varepsilon ^2 &
 6+7 \varepsilon +9 \varepsilon ^2 &
  \dots\\
 1 &
 2+2 \varepsilon +2 \varepsilon ^2 & 
 1+3 \varepsilon +5 \varepsilon ^2 &
  \\
 1 &
 0+ 2\varepsilon +2 \varepsilon ^2 &
 0- \varepsilon +\varepsilon   ^2 &
  \\
 \vdots &  &  &   \ddots \\
\end{array}
\right]
+\mathcal{O}(\varepsilon^3)
\end{align}
 The corresponding interaction vertices, sometimes called evanescent, vanish at the tree level in the limit $d\to 4$ but their presence influences loop corrections where an evanescent vertex may meet a pole resulting in a strictly finite or divergent correction. There is nothing inconsistent about the presence of evanescent interactions. The reason why they are rarely considered is that generically an evanescent part of the $\widetilde{Z}$ factors may be traded for a shift in the non-evanescent part. Hence, the space of all renormalization schemes is not made bigger (or smaller) by allowing for the evanescent interactions.

Also, we see that because the upper limit of the sum over $m$ is infinite, the same has to be true for the sum over $n$. Normally, one could as well consider the self-explanatory renormalizability contraint: $\widetilde{\lambda}^{(n,m)}=0$ for $n>2 \vee m>4$. But it does not seem easy to come up with a model where there exists a renormalization scheme that respects both the spuriosity and renormalizability constraints. In other words, we conjecture that the scale-invariant regularization is applicable exclusively to nonrenormalizable models.

\paragraph{}
The powers of $e^t$ in \eqref{thelagrangian2} and in \eqref{thelagrangian3} do not match (contrary to the powers of $\mu_0$). We claim that it does not matter. More precisely: factors $(e^t)^{\varepsilon(2n+m-2)}$ in the sum in \eqref{thelagrangian2} may be universally swapped for $(e^t)^{2\varepsilon}$ and it will not affect the resulting beta functions $\beta_{\lambda^{(n)}}$. We devote the next interim subsection to justifying this claim. Aside from that, $\mu^2=(e^t \mu_0)^2$ features explicitly in the effective action but its power there is fixed anyway by dimensionality that stems from the fields, like in $\log\frac{\phi^2}{\mu^2}$.

In summary, the generic Lagrangian \eqref{lagrangian2} describes an ordinary nonrenormalizable model. It is $\mathbb{Z}^2$-symmetric with respect to $\phi \to -\phi$ but not to $\sigma \to -\sigma$, nor does it look scale symmetric due to the nonrenormalizable part. But, if we choose
\begin{gather}
\widetilde{Z}^{(n,m)}\widetilde{\lambda}^{(n,m)} = Z^{(n)} \lambda^{(n)} a_{n,m}
\\
\quad \text{and}\quad
\mu_0=M=\sigmavev^\frac{1}{1-\varepsilon}\;,
\end{gather}
the second $\mathbb{Z}^2$ symmetry ($\sigmavev + \sigma \to - \sigmavev - \sigma$) emerges. Moreover, the model results in an effective action where the parameter $\sigmavev$ is spurious wrt the field $\sigma$. Hence, $\sigmavev+ \sigma$ is interpreted as a new field, as discussed in the previous section, and the action is seen to posses no other dimensionful parameters.

\subsection{Beta functions and the RG parameter $t$}
 
 We present here a simple algebraic calculation that supports the claim made in point 2. under Eq. \eqref{thelagrangian3}. We will examine the derivation of beta functions while allowing some freedom in how the RG-parameter $t$ enters the dimensional regularization procedure. Consider the following generic nonrenormalizable Lagrangian, $[\Phi]=1~-~\varepsilon$,
 \begin{gather}
\mathcal{L} = \frac{1}{2}Z_{\Phi}\left(\partial \Phi\right)^2 - V(\Phi)-V_\text{ct}(\Phi)
 \end{gather}
 \begin{gather}
 \begin{split}
 V + V_\text{ct} = \sum_{n=0}^\infty &\left(e^t\right)^{f(n)\varepsilon} (\mu_0)^{(n-2)\varepsilon}
 \\
 &\times Z^{(n)} \frac{\lambda^{(n)}}{n(n-1)} \frac{\Phi^n}{M^{n-4}}
\end{split} 
\end{gather}
Usually one would choose $f(n)=n-2$ and $e^t \mu_0 \equiv \mu$.

The split between $V(\Phi)$ and the counterterms $V_\text{ct}(\Phi)$ is defined by writing $Z^{(n)} = 1+\delta^{(n)}$. In minimal subtraction at one loop, the counterterms are
\begin{gather}
V_\text{ct}(\Phi) = \frac{\left[V''(\Phi)\right]^2}{64\pi^2} \frac{1}{\varepsilon} + \mathcal{O}(\varepsilon^0)
\\
V''(\Phi) = \sum_{n=2}^\infty
\lambda^{(n)} \frac{\Phi^{n-2}}{M^{n-4}} + \mathcal{O}(\varepsilon)
\\
\begin{split}
&\left[V''(\Phi)\right]^2= \sum_{n=2}^\infty\sum_{m=2}^\infty
\lambda^{(n)} \lambda^{(m)} \frac{\Phi^{n+m-4}}{M^{n+m-8}}
\\
&\quad \stackrel{ m=k-n+4}{=}
\sum_{k=0}^\infty\sum_{m=0}^{k+2} \lambda^{(k-m+4)}\lambda^{(m)} \frac{\Phi^{k}}{M^{k-4}}
\end{split}
\\
\begin{split}
&Z^{(n)} = 1 + \delta^{(n)}
\\
& = 1 + \frac{n(n-1)}{\lambda^{(n)}}
\left( \sum_{m=2}^{n+2} \lambda^{(n-m+4)}\lambda^{(m)}\right)
\frac{1}{64\pi^2} \frac{1}{\varepsilon}
\end{split}
\end{gather}
Beta function of the coupling $\lambda^{(n)}$, $\frac{\mathrm{d}\,\lambda^{(n)}}{\mathrm{d}\,t}=\beta_{\lambda^{(n)}}$, is obtained from
\begin{widetext}
\begin{gather}
0=\frac{\mathrm{d}\,}{\mathrm{d}\,t} \log \left[(e^t)^{f(n)\varepsilon}\lambda^{(n)} Z^{(n)}\right] = f(n)\varepsilon + \frac{1}{\lambda^{(n)}} \frac{\mathrm{d}\,\lambda^{(n)}}{\mathrm{d}\, t} + \frac{1}{64\pi^2} \sum_{l=0}^\infty\left[
-\lambda^{(l)} f(l)\frac{\partial}{\partial \lambda^{(l)}}
\left(
\frac{n(n-1)}{\lambda^{(n)}}\sum_{m=2}^{n+2}\lambda^{(n-m+4)}\lambda^{(m)}
\right)
\right]
\\
\begin{split}
\Rightarrow
\beta_{\lambda^{(n)}} 
&= \frac{\mathrm{d}\,\lambda^{(n)}}{\mathrm{d}\, t} = 
\frac{\lambda^{(n)}}{64 \pi^2} \sum_{l=0}^\infty \lambda^{(l)} f(l) n(n-1)\left[
\frac{-\delta_{l,n}}{{\lambda^{(l)}}^2} \sum_{m=2}^{n+2} \lambda^{(n-m-4)}\lambda^{(m)} + \frac{1}{\lambda^{(n)}} \sum_{m=2}^{n+2}\left(
\delta_{l,n-m+4}\lambda^{(m)} + \lambda^{(n-m+4)}\delta_{l,m}
\right)
\right]
\\
&= \frac{\lambda^{(n)}}{64\pi^2} n(n-1) \sum_{m=2}^{n+2}\left[
-f(n)\frac{\lambda^{(n-m+4)}\lambda^{(m)}}{\lambda^{(n)}} + f(n-m+4) \frac{\lambda^{(n-m+4)}\lambda^{(m)}}{\lambda^{(n)}} + f(m) \frac{\lambda^{(m)}\lambda^{(n-m+4)}}{\lambda^{(n)}}
\right
]\\
&=\frac{n(n-1)}{64\pi^2}\sum_{m=2}^{n+2} \frac{\lambda^{(m)}\lambda^{(n-m+4)}}{\lambda^{(n)}}\,
\underbrace{\left[ f(m) + f(n-m+4) - f(n) \right]}_{\alpha}
\end{split}
\end{gather}
 \end{widetext}
In conclusion, we have the freedom of arbitrarily choosing the function $f$ as long as the above quantity $\alpha$ is not affected. For example the ansatz $f(n) = an+b$ gives $\alpha =  4a +b$. Recall that the usual choice of $f$ is $a=1$, $b=-2$. Hence we only need to demand that $\alpha = 4-2~=~2$. Clearly the choice $a=0$, $b=2$, $f=const=2$, made in \eqref{thelagrangian3}, obeys that restriction.

\section{Scale symmetry and the role of $\mu$}
\label{roleofmuSection}

In light of the last remark in Sec. \ref{absorbingSection}, one could venture the hypothesis that the model in \eqref{thelagrangian} does not actually violate the global scale symmetry radiatively - despite beta and gamma functions being nonzero. We have proposed a construction where the derivative $\frac{\partial}{\partial t}$ is not identically equal to $\mu\frac{\partial}{\partial \mu}$ and it does not feature (at least not directly) in the Ward identity associated with the scale symmetry.
To make that more clear consider the following three dimensionality-probing operators in $d=4-2\varepsilon$ dimensions
\begin{align}
\begin{split}
\mathcal{D}_0 &\equiv (2-\varepsilon)\left[
(\partial \phi) \frac{\delta}{\delta (\partial \phi)}
+ (\partial \sigma) \frac{\delta}{\delta (\partial \sigma)}
\right]
\\
&+ (1-\varepsilon)\left[
\phi \frac{\delta}{\delta \phi} + \sigma\frac{\delta}{\delta \sigma} 
\right] - d
\end{split}
\\
\mathcal{D}_1 &\equiv \mu_0\frac{\partial}{\partial \mu_0}
\\
\mathcal{D}_2 &\equiv M \frac{\partial}{\partial M}
\end{align}
and observe that
\begin{equation}
\left(
\mathcal{D}_0 + \mathcal{D}_1 + \mathcal{D}_2
\right)\widetilde{\mathcal{L}} = 0 \;.
\label{invariance}
\end{equation}
(in this section we are treating the fields ($\phi$, $\sigma$) and the derivatives of fields ($\partial \phi$, $\partial \sigma$) as independent variables).

Normally the infinitesimal variation generated by $\mathcal{D}_0$ is associated with scale transformation, whereas $\mathcal{D}_1$ and $\mathcal{D}_2$ are not. The fact that the latter are needed in \eqref{invariance} to have zero on the RHS is regarded as a consequence of breaking of the scale symmetry. $\mathcal{D}_2$ expresses the explicit breaking by the suppression scale $M$. The breaking by $\mathcal{D}_1$ is associated with loop corrections and is usually translated into beta functions by noting that
\begin{equation}
\mu_0\frac{\partial}{\partial \mu_0} \,\widetilde{\mathcal{L}}
=
\frac{\partial}{\partial t} \, \widetilde{\mathcal{L}} = 
\sum_{n,m} \varepsilon(2n+m-2)\, \widetilde{\lambda}^{(n,m)}\frac{\partial}{\partial \vphantom{\Big|}\widetilde{\lambda}^{(n,m)}} \,
\widetilde{\mathcal{L}}
\end{equation}
So what changes when we specify from \eqref{lagrangian2} to \eqref{thelagrangian}? Nothing but the interpretation of $\mu_0$ and $M$. Observe that
\begin{gather}
\left(\mathcal{D}_1 + \mathcal{D}_2 \right) \mathcal{L} = (1-\varepsilon) \sigmavev \frac{\delta}{\delta \sigmavev} 
\,\mathcal{L}
\\
\text{and}\quad
\frac{\delta}{\delta \sigmavev} \,\mathcal{L}= \frac{\delta}{\delta \sigma}\,\mathcal{L} = \frac{\delta}{\delta (\sigmavev + \sigma)} \,\mathcal{L}
\end{gather}
Consequently
\begin{widetext}
\begin{align}
\begin{split}
&\left\{
\mathcal{D}_0 + \mathcal{D}_1 + \mathcal{D}_2
\right\} \mathcal{L}
\\
&=
\bigg\{
(2-\varepsilon)\left[
(\partial \phi) \frac{\delta}{\delta (\partial \phi)}
+ (\partial (\sigmavev+\sigma)) \frac{\delta}{\delta (\partial (\sigmavev+\sigma))}
\right]
+ (1-\varepsilon)\left[
\phi \frac{\delta}{\delta \phi} + (\sigmavev+\sigma)\frac{\delta}{\delta (\sigmavev+\sigma)} 
\right] - d
\bigg\}\,\mathcal{L} \,=\, 0
\end{split}
\label{invariance2}
\end{align}
\end{widetext}
and now the whole operator in curly braces is interpreted as the generator of dilatations.

To be clear, the above Eqs. (\ref{invariance}-\ref{invariance2}) deal only with the classical Lagrangians $\mathcal{L}$ and $\widetilde{\mathcal{L}}$, albeit in $d$ dimensions. In order to convincingly talk about scale symmetry at the quantum level and the vanishing of radiative scale symmetry breaking, one should properly rederive Ward-Takahashi identities generated by scale transformations for the Green functions and we devote the next half of the report to this task. Crucially though, it is before such an analysis is undertaken that one has to define how the symmetry is represented by transformations of various quantities, i.e. fields and parameters. Scale invariant regularization offers a new perspective on this prerequisite step and instructs us to transform the renormalization scale as if it were a field background. One possibility would be to do that in the course of spurion analysis, to simply proclaim that $\mu$ scales like dimension 1 operator, and call it a day. The more interesting alternative discussed in this report is to arrange for the renormalization scale to be consistently absorbed by a background of a genuine dynamical field.

\section{Quantum scale invariance}
\label{Quantum-scale-invariance-section}

In the framework of a quantum field theory we are interested in scaling as a transformation that acts on renormalized Green functions $G^N$,

\begin{align}
& G^N_{0}(x_1,\ldots,x_n) = \int \mathcal{D} \phi_0 \, \phi_0(x_1)\ldots \phi_0(x_N)\,e^{i\int \mathcal{L}[\phi_0]}
\label{pathintegral}
\\
\label{greenfunction}
& \equiv Z_\phi^{\frac{N}{2}} G^N(x_1, \ldots, x_N)
\;,\quad 
\phi_0 \equiv Z_\phi^{\frac{1}{2}} \phi \;.
\end{align}
(We suppress normalization factors in front of path integrals, like in the one in \eqref{pathintegral}, throughout the report. By definition, background fields $\phivev$, $\sigmavev$ discussed in section \ref{spuriositySection} do not enter those normalization factors.)

We will be using  MS scheme throughout and assuming applicability of a perturbative expansion in couplings, generically denoted with $\{\lambda\}$. The renormalization factors $Z_\#$, including $Z_\phi$, implicitly contain the following perturbative structure in $d = 4 - 2 \varepsilon$ dimensions
\begin{equation}
Z_\# = 1+ \frac{f^\#_1}{2 \varepsilon} + \frac{f^\#_2}{(2 \varepsilon)^2} + \ldots\;,\quad f^\#_m = \mathcal{O}(\{\lambda\}^m)\;.
\label{Zfactors}
\end{equation} 
By an (infinitesimal) scale transformation of a finite, physically relevant function that is $G^N$, we mean
\begin{equation}
\left(Nd_\phi+ \sum_{i=1}^N x_i^\mu \frac{\partial}{\partial x^\mu_i} \right) G^N(x_1, \ldots, x_N) \stackrel{?}{=} 0\;.
\label{quantumscaling}
\end{equation}
If the classical Lagrangian $\mathcal{L}$ is free of dimensionfull parameters, the above equality is satisfied at the tree level, but typically it is invalidated by loop corrections. The RHS is given in a precise nonperturbative form in terms of beta functions and anomalous field dimensions. This way, by considering scale transformations at the quantum level, one circles back to the Callan-Symanzik equation for $G^N$ that expresses its independence of renormalization scheme. We will present this in detail below.
On the other hand, the RHS of \eqref{quantumscaling} being exactly zero to all orders defines what we will refer to as \textit{quantum scale invariance}.

We claim that quantum scale invariance follows from the scale invariant regularization introduced in the previous sections. This is despite the fact that a nontrivial Callan-Symanzik equation is also satisfied by a quantum scale invariant theory.

The hypothetical equality in \eqref{quantumscaling} can be equivalently expressed for other types of Green functions and in the momentum rather than position space. The resulting expressions are not hard to find. An $N$-particle 1PI function in momentum space has dimension $(d-Nd_\phi)$, and so, given \eqref{quantumscaling}, it satisfies
\begin{equation}
\left[(d - N d_\phi) - \sum_{i=1}^{N} p_i \frac{\partial}{\partial p_i} \right] \Gamma^N (p_1, \ldots,p_N)= 0\;.
\end{equation}
Recall also that the effective potential is given by the sum of momentum-independent 1PI diagrams, for which
\begin{equation}
0 = \left( d - N d_\phi \right) \Gamma^N_{p=0} = \left(d - d_\phi \phi \frac{\partial}{\partial \phi} \right)\Gamma^N_{p=0}\;.
\end{equation}
In conclusion, if follows from \eqref{quantumscaling} that the effective potential in quantum scale invariant model is a homogeneous function of fields,
\begin{equation}
\left(d - d_\phi \phi \frac{\partial}{\partial \phi}\right) V_\text{eff} \, =\, 0.
\end{equation}
This, let us stress again, is to be understood as a nonperturbative statement.

\section{Symmetry transformation as an operator insertion}

The instrumental fact for our analysis is that the scale transformation \eqref{quantumscaling} is equivalent to an insertion of a renormalized composite operator. For a valuable introduction to the notion of inserting an operator into Green functions we refer the Reader to \cite{ticciati1999quantum}.
Much of the calculations in this and in the next section can be characterized as an exercise in algebraic renormalization and in using the so called quantum action principle. As our goal is to present explicit but concise line of argumentation, we do not review those topics here and only refer the Reader to \cite{Breitenlohner:1977hr} for original development and to \cite{Piguet:1995er,Franco:2013} for a more recent presentation and further references.

Recall that the Green function $G^N$ in \eqref{greenfunction} is properly renormalized only when $\forall_{i\neq j} x_i \neq x_j$. Considering e.g. the limit $x_1,\,x_2 \to x$ leads one to the notion of composite operator $\phi_0(x_1) \phi_0(x_2) \to \phi_0^2(x)$ inserted into $G_0^{N-2}$ resulting in
\begin{align}
\begin{split}
&G_0^{N-2}\left(x_1, \ldots, x_{N-2};\phi_0^2(x)\right)
\\
& \equiv 
\int \!\mathcal{D} \phi_0 \;\phi_0^2(x)\,\phi_0(x_1)\ldots\phi_0(x_{N-2}) \, e^{i\int\! \mathcal{L}[\phi_0]}.
\end{split}
\end{align}
We are showcasing here the notation that will be used henceforth for insertion of any operator $\mathcal{O}_0$,
\begin{align}
\begin{split}
&G_0(\mathcal{O}_0(x)) = G_0\left(x_1, \ldots; \mathcal{O}_0(x)\right)
\\
&= 
\int \!\mathcal{D} \phi_0 \;\mathcal{O}_0(x)\,\phi_0(x_1)\ldots e^{i\int\! \mathcal{L}}.
\end{split}
\end{align}
The factors of $Z_\phi$ alone are generically not sufficient to renormalize bare Green function $G_0$ with an operator insertion. But the operator itself can be renormalized, $\mathcal{O}_0(x) = Z_\mathcal{O} \left[\mathcal{O} \right](x)$, so that
\begin{equation}
G^N\left(\left[\mathcal{O}\right]\right) = Z_\mathcal{O}^{-1} Z_\phi^{-\frac{N}{2}}G_0^N(\mathcal{O}_0)
\end{equation}
is a finite function for any $N$. We will denote the feature of being renormalized by enclosing an operator in square brackets.

The situation is actually more complicated by the fact that renormalization usually requires mixing of operators,
\begin{align}
\begin{split}
\mathcal{O}_0(x) = \sum_{\mathcal{O}} Z_\mathcal{O}\left[ \mathcal{O}\right]\;,\quad  \text{where the sum is}
\\
\text{over $\left[\mathcal{O}\right]$ such that} \quad
\text{dim}\left[ \mathcal{O} \right]\leqslant \text{dim}\, \mathcal{O}_0\,.
\label{operatorexpansion}
\end{split}
\end{align}

Importantly, there is a class of composite operators of the form $\left(  \int\!\mathrm{d}^dx \,\delta\phi_0\frac{\delta \mathcal{L}}{\delta \phi_0}\right)$ which do not require renormalization with divergent $Z$-factors. Consider an unspecified field variation $\phi_0 \to \phi_0 + \epsilon \delta \phi_0$ and the corresponding infinitesimal variation in the dummy variable of path integration,
\begin{widetext}
\begin{gather}
e^{i Z[J]} = \int\!\mathcal{D}\phi_0\,e^{i\int\left( \mathcal{L}[\phi_0] + \phi_0 J\right)}
= \int\!\mathcal{D}(\phi_0 + \epsilon \delta \phi_0) \,e^{i\int\left( \mathcal{L}[\phi_0+ \epsilon \delta \phi_0] + (\phi_0 + \epsilon \delta \phi_0) J\right)} \\
= e^{iZ[J]} +  i \epsilon\int\!\mathcal{D}\phi_0\left( \delta \phi_0 \frac{\delta \mathcal{L}}{\delta \phi_0} + \delta\phi_0 J \right) e^{i\int \left(\mathcal{L}[\phi_0] + \phi_0 J\right)} + \mathcal{O}(\epsilon^2) \\
\Rightarrow\;
\int\!\mathcal{D}\phi_0\;\int\!\mathrm{d}^dx\!\left(\! -i\delta \phi_0(x) \frac{\delta \mathcal{L}}{\delta \phi_0}[\phi_0(x)] \right)\;
e^{i\int(\mathcal{L} + \phi_0 J)}
=
\int\!\mathcal{D}\phi_0\;\int\!\mathrm{d}^dx\!\left(i\delta\phi_0(x) J(x) \right)\;
e^{i\int(\mathcal{L} + \phi_0 J)}
\label{dummyvariablechange}
\end{gather}
\end{widetext}
We have assumed here that the integration measure is invariant with respect to the unspecified transformation, $\mathcal{D}\phi_0 \!=\! \mathcal{D}(\phi_0 \!+\! \epsilon\delta\phi_0)$. This is not always true. Transformations that produce a nontrivial Jacobian are famously related to quantum anomalies, radiative violations of symmetries albeit of a different kind than the one we are discussing here. If we were to consider a model in curved spacetime, where the spacetime dilatations necessarily have to be regarded as local transformations, there would in fact be a nontrivial jacobian given in terms of Riemann and Weyl tensors, resulting in what is known as Weyl- (or conformal-) anomalies \cite{fujikawa2004path}.

From \eqref{dummyvariablechange}, after applying $\left(\frac{-i\,\delta}{\delta J(x_1)}\right) \!\ldots\! \left(\frac{-i\,\delta}{\delta J(x_N)}\right)\ast~\bigg|_{J=0}$, to both sides, and stripping away the timespace integration, it follows that
\begin{gather}
\begin{split}
&G_0^N\left(x_1, \ldots,x_N;-i\left(\delta\phi_0 \frac{\delta\mathcal{L}}{\delta \phi_0}\right)(x)\right) 
\\
&= \sum_{i=1}^N \delta(x_i-x)\, G_0^N(x_1,\ldots,\hat{x}_i,\ldots,x_N)\;,
\\
&\text{where}\quad G_0^N(x_1,\ldots,\hat{x}_i,\ldots,x_N)
\\
&\equiv \int\!\mathcal{D}\phi_0\,\phi_0(x_1)\ldots\delta\phi_0(x_i)\ldots\phi_0(x_N)\,e^{i\int\!\mathcal{L}[\phi_0]}
\end{split}
\end{gather}
Now we specify $\delta \phi_0$ to one one of the two following possibilities and define the corresponding composite operator, $E_0$ or ${E_\mu}_0$ below.
\begin{itemize}
\item $\delta\phi_0(x) \equiv \phi_0(x)$
\begin{gather}
\begin{split}
&G_0^N \Big( x_1,\ldots,x_N;\underbrace{-i\left(\phi_0 \frac{\delta\mathcal{L}}{\delta \phi_0}\right)(x)}_{\equiv\, i\,E_0(x)} \Big)
\\
&= \sum_{i=1}^N\delta(x_i-x)\,G_0^N(x_1,\ldots,x_N)
\label{Edefinition}
\end{split}
\end{gather}
\item $\delta\phi_0(x) \equiv \partial_\mu \phi_0(x)$
\begin{gather}
\begin{split}
G_0^N \Big( x_1,\ldots,x_N;\underbrace{-i\left(\partial_\mu\phi_0 \frac{\delta\mathcal{L}}{\delta \phi_0}\right)(x)}_{\equiv\, i\,{E_\mu}_0(x)} \Big)
\\
= \sum_{i=1}^N\delta(x_i-x)\,\frac{\partial}{\partial x^\mu_i}G_0^N(x_1,\ldots,x_N)
\label{Emudefinition}
\end{split}
\end{gather}
\end{itemize}
Since the RHS of \eqref{Edefinition} and \eqref{Emudefinition} can be renormalized by the $Z_\phi^{-\frac{N}{2}}$ factors alone, the same is true about the LHS. In other words, in the MS scheme,
\begin{gather}
Z_{E} = Z_{E_\mu} = 1 \;,\quad E_0(x) = [E](x)\;,\quad {E_\mu}_0(x) = [E_\mu](x)\;.
\end{gather} 
Note that one can use insertion of $[E]$ to multiply $G^N$ by the number $N$
\begin{equation}
G^N\left(i\int\!\mathrm{d}^dx[E](x)\right) = N G^N\;.
\label{NfactorbyEinsertion}
\end{equation}
Now we can finally write down the renormalized composite operator $[\mathcal{O}^S]$, whose insertion generates scale transformation \eqref{quantumscaling} in the following sense
\begin{align}
&[\mathcal{O}^S](x) \, \equiv\, id_\phi [E](x) + i x^\mu [E_\mu](x)
\\
\begin{split}
&G^N\left( x_1, \ldots, x_N ; [\mathcal{O}^S](x) \right) 
\\
&=\sum_{i=1}^N\delta(x_i-x)\,\left(d_\phi + x_i^\mu \frac{\partial}{\partial x^\mu_i} \right) G^N(x_1,\ldots,x_N)
\end{split}
\\
\begin{split}
&G^N\left( x_1, \ldots, x_N ; \int\!\mathrm{d}^dx[\mathcal{O}^S](x) \right) 
\\
&=\left( Nd_\phi + \sum_{i=1}^N x_i^\mu \frac{\partial}{\partial x^\mu_i} \right) G^N(x_1,\ldots,x_N)
\end{split}
\label{OSinsertion}
\end{align}
Thus, our definition \eqref{quantumscaling} of quantum scale invariance can be restated in terms of invariance with respect to the insertion of $\int\!\mathrm{d}^dx[\mathcal{O}^S]$.  This brings us directly to the energy-momentum tensor.

\section{Scale symmetry violation and loop corrections}
\label{loopsSection}

The reason why it is convenient to express a symmetry transformation as an operator insertion is that we can relate $[\mathcal{O}^S](x)$ to another operator that is a full four-divergence, in the following way
\begin{equation}
[\mathcal{O}^S](x) \,+\,  [\mathcal{O}^B](x) \;=\; i \partial_\mu [D^\mu](x)
\label{OSB}
\end{equation}
and so it follows that
\begin{equation}
G\left(\int\!\mathrm{d}^dx[\mathcal{O}^S]\right) = -G\left(\int\!\mathrm{d}^dx[\mathcal{O}^B]\right)\;,
\label{OSOB}
\end{equation}
because we will be assuming that $G([D^\mu])$ vanishes at the boundary at infinity for any Green function. The current $[D^\mu]$ is a renormalized operator, given at tree level by the classical Noether current that corresponds to the symmetry under consideration. We are assuming that the classical Lagrangian is scale invariant and so we are prompted to use the dilatation current
\begin{gather}
[D^\mu] =  x_\nu[T^{\mu\nu}]
\\
\text{whence}\quad \partial_\mu [D^\mu] = [T^\mu_{\;\;\mu}] + x_\nu\partial_\mu [T^{\mu\nu}]\;,
\end{gather}
with the energy-momentum tensor operator $[T^{\mu\nu}]$, in terms of the bare field, 
\begin{align}
&[T^{\mu\nu}] =
\label{EMtensor}
\\
\nonumber
& \frac{\delta \mathcal{L}[\phi_0]}{\delta\partial_\mu \phi_0} \partial^\nu \phi_0 - g^{\mu\nu} \mathcal{L}[\phi_0] - \frac{d-2}{4(d-1)}\left(\partial_\mu\partial_\nu - g_{\mu\nu}\partial^2\right)\phi_0^2
\end{align}
Despite appearances, thanks to the fact that $\mathcal{L}$ is scale symmetric, if we choose $\Delta\eta = 0$ (see Appendix \ref{appendix}), $[T^{\mu\nu}]$ does not require any divergent counterterms to be renormalized.

For the purpose of our discussion we can simply regard $[D^\mu]$ as a promising ansatz and \eqref{OSB} as the definition of $[\mathcal{O}^B]$. Our goal now is to obtain the explicit form of this equation and the operator $[\mathcal{O}^B]$ in particular, and analyze the validity or breaking of quantum scale invariance that is encoded in \eqref{OSOB}. First we will use the general approach to dimensional regularization with the regular inert renormalization scale $\mu$. Second, we will recast the calculation for the model presented in \eqref{thelagrangian} that is applying the scale invariant regularization. Our analysis will closely follow (in the first part) or be inspired by (in the second part) the article \cite{Brown:1979pq}. (Especially in the context of non-scale-invariant regularization the Reader is referred there for a more detailed and precise presentation that also treats an explicit scale symmetry violation by a mass term.)

\subsection{Regular approach}
The simplest classically scale-symmetric example one can analyze is the $\lambda \phi^4$ model, (node in the lower index denotes a bare field or coupling)
\begin{equation}
\mathcal{L}[\phi_0] = \frac{1}{2} \left(\partial \phi_0\right)^2 - \frac{\lambda_0}{4} \phi_0^4\;.
\label{lambdaphitothefourth}
\end{equation}
Using formulas \eqref{Edefinition}, \eqref{Emudefinition}, \eqref{EMtensor}, in light of our discussion so far, we have
\begin{gather}
\nonumber
[E] = \phi_0 \partial^2 \phi_0 + \lambda_0 \phi_0^4
\;, \quad
[E_\mu] = \left( \partial_\mu \phi_0\right)\left(\partial^2 \phi_0 + \lambda_0 \phi_0^3\right) ,
\\
\begin{split}
[T^\mu_{\;\;\mu}] &= (\partial \phi_0)^2 - \frac{d}{2}(\partial \phi_0)^2 + d \frac{\lambda_0}{4} \phi_0^4 + \frac{(d-2)}{4}(\partial^2 \phi_0^2)
\\
&=\left(\frac{d}{2} - 1\right)\left([E] - \lambda_0 \phi_0^4\right) + d \frac{\lambda_0}{4} \phi_0^4
\\
&= d_\phi [E] + 2\varepsilon \frac{\lambda_0}{4} \phi_0^4\;.
\label{TmumuReg}
\end{split}
\end{gather}
Notice the "evanescent" contribution at the end of the last expression. Its presence is crucial for our analysis, as it is the very source of scale symmetry violation by loop corrections that we are interested in. Going further,
\begin{align}
\begin{split}
\partial_\mu[T^{\mu\nu}]
&=\left(\partial_\mu \phi_0\right)\left(\partial^\mu \partial^\nu \phi_0\right) + \left(\partial^\nu \phi_0\right)\left(\partial^2 \phi_0\right)
\\
&\qquad - \left(\partial^\nu\partial^\mu \right)\left(\partial_\mu \phi_0\right) + \lambda_0 \phi_0^3 \partial^\nu \phi_0
\\
&= [E^\nu]
\end{split}
\label{dmuTmunuReg}
\end{align}
and in the end
\begin{gather}
i\,\partial_\mu[D^\mu] = i\, d_\phi[E] +i\,x^\nu [E_\nu] + 2i\, \varepsilon\frac{\lambda_0}{4} \phi_0^4 \\
[\mathcal{O}^B](x) = 2i\,\varepsilon\frac{\lambda_0}{4} \phi_0^4(x) \neq 0 
\label{OBfirst}
\end{gather}
It is not yet evident that the RHS of \eqref{OBfirst} is a finite operator in the $\varepsilon \to 0$ limit, as we expect, or how to interpret its insertion into a renormalized Green function. Valuable insight is gained by performing renormalization of the $\phi_0^4$ composite operator.

\paragraph{Renormalization of the $\phi_0^4$ operator}

As discussed earlier along Eq. \eqref{operatorexpansion}, on general grounds we have
\begin{equation}
\frac{\lambda_0}{4} \phi_0^4 = Z^1_{\phi^4} \mu^{2\varepsilon} \frac{\lambda}{4} [\phi^4] + 
(Z^2_{\phi^4} - 1)[E] + (Z^3_{\phi^4} - 1)\partial^2 [\phi^2]
\label{phi4expansion}
\end{equation}
which is a combination of all $(4-2\varepsilon)$-dimensional  operators at our disposal (e.g. $[(\partial\phi)^2]$ is already linearly dependent). Recast the above in the following way
\begin{gather}
Z_{\phi4}^{i} \equiv 1 + \frac{a^{(i)}}{2\varepsilon}\;,\quad
a^{(i)} = a^{(i)}(\varepsilon)\;,\quad
i=1,2,3\;, \quad\text{so that}
\nonumber
\\
\frac{\lambda_0}{4}\phi_0^4 = \left(1 + \frac{a^{(i)}}{2\varepsilon} \right)\mu^{2 \varepsilon} \frac{\lambda}{4} [\phi^4] + \frac{a^{(2)}}{2 \varepsilon} [E] + \frac{a^{(3)}}{2 \varepsilon} \partial^2 [\phi^2]
\label{phi4ansatz}
\end{gather}
The Author of \cite{Brown:1979pq} presents an extremely convenient way of finding $a^{(1)}$, $a^{(2)}$ and $a^{(3)}$ by relating $\frac{\partial}{\partial \lambda_0}$ and $\frac{\partial}{\partial \lambda}$ acting on a renormalized $G^N$ function. He concluded that
\begin{align}
\frac{\partial}{\partial \lambda} \ast =
&
i\,\int\!\mathrm{d}^d x \bigg[
\frac{2\varepsilon + a^{(1)}}{\beta-2 \varepsilon \lambda} \mu^{2 \varepsilon} \frac{\lambda}{4}[\phi^4](x)
\\
\nonumber
&+ \frac{-\gamma_\phi + a^{(2)}}{\beta-2 \varepsilon \lambda} [E](x) + \frac{a^{(3)}}{\beta-2 \varepsilon \lambda} \partial^2 [\phi^2](x) \bigg] \ast
\label{ast}
\end{align}
where $\beta$ and $\gamma$ have the standard meaning,
\begin{gather}
\begin{split}
\lambda_0 &= \left(e^t\right)^{2 \varepsilon} \lambda Z_\lambda\\
0 \stackrel{!}{=} \frac{\mathrm{d}}{\mathrm{d}t} \lambda_0
&\Rightarrow \frac{\partial \lambda}{\partial t} = -2 \varepsilon \lambda + \beta(\lambda)
\end{split}
\\
\phi_0 = Z_\phi^{\frac{1}{2}}\phi
\;,\quad
\frac{\mathrm{d}}{\mathrm{d}t} Z_\phi^{\frac{1}{2}} = 
Z_\phi^{\frac{1}{2}} \gamma_\phi
\end{gather}
From finiteness of $\frac{\partial}{\partial \lambda}G^N$  it follows that
\begin{gather}
a^{(1)} = -\frac{\beta}{\lambda}\;,\quad
a^{(2)} = \gamma_\phi \;,\quad
a^{(3)} = 0
\end{gather}
One concludes that, in the MS scheme,
\begin{equation}
\frac{\partial}{\partial \lambda} G^N= G^N\!\left(-
\int\!\mathrm{d}^d x\,\mu^{2 \varepsilon}\frac{i}{4}[\phi^4](x)\right)
\end{equation}
and that \eqref{phi4expansion} takes the explicit form
\begin{equation}
\frac{\lambda_0}{4} \phi_0^4 = \left(1-\frac{\beta/\lambda}{2\varepsilon}\right)\frac{\lambda}{4}\mu^{2\varepsilon}[\phi^4] + \frac{\gamma_\phi}{2\varepsilon}[E]
\label{phi4renormalizationEND}
\end{equation}

Coming back to the scale transformation and to the formula \eqref{OBfirst}, we finally have the $[\mathcal{O}^B]$ operator explicitly regularized and a handy interpretation along with it,
\begin{gather}
[\mathcal{O}^B]  = (2\varepsilon \lambda - \beta)\frac{i}{4} \mu^{2\varepsilon} [\phi^4] + i \gamma_\phi [E]\\
\left(\int\!\mathrm{d}^d x\,[\mathcal{O}^B](x)\right)\ast 
\,=\, 
\left[
(\beta - 2\varepsilon \lambda )\frac{\partial}{\partial \lambda} + \gamma_\phi N\right]\ast
\end{gather}
so that, using \eqref{OSinsertion} and \eqref{OSOB}, our original query \eqref{quantumscaling} about the quantum scaling can finally be answered.
\begin{align}
\begin{split}
&\left(Nd_\phi + \sum_{i=1}^N x_i^\mu \frac{\partial}{\partial x_i^\mu}\right)
G^N(x_1,\ldots,x_N) 
 \\
 &
 \hspace{0cm}
= - \left[
(\beta - 2\varepsilon \lambda)\frac{\partial}{\partial \lambda} + \gamma_\phi N\right]G^N 
 \\
  &\mathrel{\overset{\varepsilon \to 0}{\longrightarrow}}\;
 -\left(\beta\frac{\partial}{\partial\lambda} + \gamma_\phi N\right)G^N
 \end{split}
 \label{regularscaling}
\end{align}
Thus, our notion of quantum scale invariance is clearly not realized in this example. Furthermore, one can couple the above equation with dimensional analysis in the well known fashion. Function $G^N$ itself has dimension $Nd_\phi$ and all of its dimensionful arguments are presumably $x_1, \ldots, x_N$ and $\mu_0$. Hence, for any $s \in \mathbb{R}$,
\begin{gather}
\begin{split}
&G^N=G^N(x_1, \ldots, x_N; \mu_0) = s^{Nd_\phi} G^N\!\!\left(sx_1,\ldots, sx_N; \frac{\mu_0}{s}\right)
\\
&s\frac{\partial}{\partial s}G^N = \left(Nd_\phi
 + \sum_{i=1}^N x_i^\mu \frac{\partial}{\partial x_i^\mu} - \mu_0 \frac{\partial}{\partial \mu_0}
 \right)G^N = 0
\end{split}
 \label{dimensionalanalysis}
\end{gather}
If we additionally posit that $\mu_0\frac{\partial}{\partial \mu_0} = \frac{\partial}{\partial t}$, \eqref{regularscaling} and \eqref{dimensionalanalysis} combine to the Callan-Symanzik equation for $G^N$,
\begin{equation}
\left(\frac{\partial}{\partial t} + \beta \frac{\partial}{\partial \lambda} + N\gamma_\phi\right)G^N = 0
\label{CallanS}
\end{equation}
Let us stress, in a way of summary, that the Eq. \eqref{regularscaling} is an operator equation in disguise and that it is this relation that captures the behavior of Green functions under scale transformations. The reference to $\mu_0$ is made in the very last part of our discussion in the context of elementary dimensional analysis when the $\mu_0$-dependence is traded for $t$-dependence to arrive at \eqref{CallanS}. The quick jump from \eqref{regularscaling} to \eqref{CallanS}, from the language of operator insertions to renormalization group, is very elegant and suggestive. The nontrivial running of couplings and anomalous field dimensions are widely interpreted as a robust manifestation of how the inherent need to renormalize a quantum field theory translates into anomalous, i.e. loop-generated, breaking of the scale symmetry. In the next section we are challenging this connection by considering a model regularized in the scale invariant way.

\subsection{Example with the scale invariant regularization}
We aim to repeat the program of finding the renormalized composite operator $[\mathcal{O}^B]$, this time for the model in \eqref{thelagrangian},
\begin{gather}
\mathcal{L}[\phi_0,\sigma_0] =
\frac{1}{2}\left(\partial \phi_0\right)^2 + 
\frac{1}{2}\left(\partial \sigma_0\right)^2 - V 
\\
V = \left[e^t\sigma_0^{\frac{1}{1-\varepsilon}}\right]
\sum_{n=0}^\infty \frac{\lambda_0^{(n)}}{4}\phi_0^{2n}\sigma_0^{4-2n}
\\
e^{iZ[J\phi,J_\sigma]} = \int\!\mathcal{D}\phi_0 \mathcal{D}\sigma_0\,e^{i\int\!\left(\mathcal{L}[\phi_0,\sigma_0] + \phi_0 J_\phi + \sigma_0 J_\sigma\right)}
\end{gather}
We also define the $[E]$, $[E_\mu]$, and $[\mathcal{O}^S]$ operators in a complete analogy to the previous case,
\begin{gather}
\begin{split}
&[E^\phi] = -\phi_0 \frac{\delta}{\delta \phi_0} \mathcal{L} = \phi_0 \partial^2\phi_0 + 2nV
\\
&[E^\sigma] = -\sigma_0 \frac{\delta}{\delta \sigma_0} \mathcal{L} = \sigma_0 \partial^2 \sigma_0 + \left(4-2n + \frac{2\varepsilon}{1-\varepsilon}\right)V
\end{split}
\\
\begin{split}
[E] &\equiv [E^\phi] + [E^\sigma]
\\
&= 
\phi_0 \partial^2\phi_0 +  \sigma_0 \partial^2 \sigma_0 + \frac{d}{d_\phi} V 
\;,\quad d_\phi = d_\sigma
\end{split}
\\
\begin{split}
&[E^\phi_\mu] = -(\partial_\mu\phi_0)\frac{\delta}{\delta \phi_0} \mathcal{L}
\;,\quad
[E^\sigma_\mu] = -(\partial_\mu\sigma_0)\frac{\delta}{\delta \sigma_0} \mathcal{L}
\\
&[E_\mu] \equiv [E_\mu^\phi] + [E_\mu^\sigma] = 
\partial_\mu(\partial^2\phi_0) + \partial_\mu(\partial^2\sigma_0) + \partial_\mu V
\end{split}
\end{gather}
\begin{gather}
[\mathcal{O}^S] = i[E] + ix^\mu [E_\mu]
\end{gather}
The new feature afforded to us by the scale invariant form of $V$ is the factor of $d/d_\phi$ rather than 4 in front of $V$ in the $[E]$ operator.
Further,
\begin{widetext}
\begin{gather}
\begin{split}
[T^\mu_{\;\;\mu}]
 &= (\partial \phi_0)^2 + 
  (\partial \sigma_0)^2 - d\left[\frac{1}{2}\left(\partial \phi_0\right)^2 + \frac{1}{2}\left(\partial \sigma_0\right)^2 - V\right] + \frac{d-2}{4}\partial^2\left(\phi_0^2 + \sigma_0^2\right) 
\\
 &= \left(\frac{d}{2} - 1\right)\left(\phi_0\partial^2\phi_0 + \sigma_0 \partial^2\sigma_0\right) + dV + \left[\frac{1}{2}\left(1-\frac{d}{2}\right) + \frac{d-2}{4} \right] \partial^2\left(\phi_0^2 + \sigma_0^2\right)
\\
  &= d_\phi\left([E] - \frac{d}{d_\phi}V\right) + d V = d_\phi [E]
\end{split}
\label{TmumuScaleInvariant}
\end{gather}
and as before
\begin{align}
\begin{split}
\partial_\mu[T^{\mu\nu}] = &
(\partial_\mu\phi_0)(\partial^\mu \partial^\nu \phi_0) +
(\partial_\mu\sigma_0)(\partial^\mu \partial^\nu \sigma_0)+
(\partial^\nu\phi_0)(\partial^2\phi_0) + 
(\partial^\nu\sigma_0)(\partial^2\sigma_0) 
\\
&\hspace{3.4cm}
- (\partial_\nu \partial_\mu \phi_0)(\partial^\mu \phi_0)
- (\partial_\nu\partial_\mu \sigma_0)(\partial^\mu \sigma_0) + \partial^\nu V \, = \, [E^\nu]
\end{split}
\label{dmuTmunu}
\end{align}
\end{widetext}
Thus we reach the pleasantly simple result
\begin{align}
\begin{split}
[\mathcal{O}^S] &= i d_\phi [E] + i x^\mu[E_\mu] = i \partial_\mu\left(x_\nu[T^{\mu\nu}]\right) = i\partial_\mu D^\mu \\
[\mathcal{O}^B] &= 0
\end{split}
\label{OBzero}
\end{align}
and the quantum scale invariance, as defined by \eqref{quantumscaling}, is realized.

A general Green function in this simple model depends on two sets of variables that correspond respectively to its $\phi$- and $\sigma$-legs,
\begin{gather}
\begin{split}
&G^{N,M}(x_1,\ldots,x_N;y_1,\ldots,y_M)
\\
&= \int\!\mathcal{D}\phi\mathcal{D}\sigma\,\phi(x_1)\ldots\phi(x_N)\,\sigma(y_1)\ldots\sigma(y_M)\,e^{i\!\int\!\mathcal{L}}
\end{split}
\end{gather}
In light of \eqref{OBzero}, $G^{N,M}$ satisfies
\begin{equation}
\left[
(N+M)d_\phi + \sum_{i=1}^N x^\mu_i\frac{\partial}{\partial x^\mu_i} + \sum_{j=1}^M y_j^\mu \frac{\partial}{\partial y_j^\mu}
\right]
G^{N,M}  = 0
\label{quantumscaleinvarianceExample}
\end{equation}
and we conclude that the effective potential is scale-invariant to all orders,
\begin{equation}
\left(d - d_\phi \phi\frac{\partial}{\partial \phi} - d_\phi\sigma\frac{\partial}{\partial \sigma}\right)V_\text{eff}(\phi,\sigma) \,=\, 0\,.
\label{Vhomogeneous}
\end{equation}

In a quantum scale invariant model, the only possible minima of the effective potential $V_\text{eff}$ that is bounded below (and thus likely the vacua of the model) are either a global zero-dimensional minimum at $\Phi=0$ (where $\Phi$ stands for all scalar fields) or an (at least) one-dimensional flat direction with a continuum of degenerate minima. Furthermore, only the second option allows for spontaneous breaking of the scale symmetry, which we are assumming takes place. In other words the quantum scale invariant model can either be in the completely stable unbroken phase or be inconsistent due to the unboundedness below of the effective potential. The infinitesimal boundary between those two possibilities is the only region in the parameter space that we hope can be deemed physically relevant. There the scale symmetry is spontaneously broken and the potential is still bounded. How or why should the theory find itself at that boundary, we do not explain.

It is also illuminating to consider renormalization of the composite operator given by the potential, $[V]$. This is done in much the same fashion as for $[\phi^4]$ in the previous example, see Eqs. (\ref{phi4expansion}-\ref{phi4renormalizationEND}), despite the fact that the potential now contains an infinite series of interactions, each with an independent coupling $\lambda^{(i)} \in \{\lambda\}$, whose $\beta^{(i)}$ function again depends on infinitely many couplings,
\begin{gather}
\lambda_0^{(i)} = \left(e^t\right)^{2\varepsilon}Z_{\lambda^{(i)}}\lambda^{(i)}\;,\quad Z_{\lambda^{(i)}}\equiv e^{f^{(i)}}\\
\beta^{(i)}=\beta^{(i)}(\{\lambda\}) = \lambda^{(i)}\sum_{j} \frac{\partial f^{(i)}}{\partial \lambda^{(j)}}\left(2\varepsilon\lambda^{(j)} - \beta^{(j)}\right)
\end{gather}
Consequently, in analogy to \eqref{phi4renormalizationEND}, we arrive at
\begin{gather}
\begin{split}
&\sum_{i}\left(2\varepsilon\lambda^{(i)} - \beta^{(i)}\right)\frac{\partial}{\partial\lambda^{(i)}} G^{N,M} 
\\
&= 
\left(N \gamma_\phi + M \gamma_\sigma + 2\varepsilon\sum_{j} \lambda_0^{(j)}\frac{\partial}{\partial \lambda_0^{(j)}}\right)
G^{N,M}
\end{split}
\label{2epslambetadGdlam}
\end{gather}
We can rewrite the last equation to have the potential (as the bare operator) on the left-hand side,
\begin{gather}
\begin{split}
\left(\int\!\mathrm{d}^dx\,V(x)\right)\ast
=
& \sum_i\left(1 - \frac{\beta^{(i)}/\lambda^{(i)}}{2\varepsilon}\right) i\lambda^{(i)}\frac{\partial}{\partial \lambda^{(i)}} \ast \;\,
\\
&+\int\!\mathrm{d}^dx\left(
\frac{\gamma_\phi}{2\varepsilon} [E_\phi] + \frac{\gamma_\sigma}{2\varepsilon} [E_\sigma]
\right)\ast 
\end{split}
\end{gather}
Clearly the $\beta$ and $\gamma$ functions are still needed to renormalize the potential. Similarly, we still expect the Green functions to be independent of the regularization parameter $t$, and thus to satisfy the Callan-Symanzik equation
\begin{equation}
\left(\frac{\partial}{\partial t} + \sum_{i}\beta^{(i)} \frac{\partial}{\partial \lambda^{(i)}} + N\gamma_\phi + M\gamma_\sigma\right)G^{N,M} = 0\,.
\label{CallanS2}
\end{equation}
This has been checked explicitly in the unusual scheme [defined by adhering to the MS principle for the coefficients in \eqref{anm}] at the level of two loops for the effective potential in the discussed model in \cite{Ghilencea:2016ckm}, see also \cite{Ghilencea:2017yqv}.

Notice that there is no contradiction between \eqref{quantumscaleinvarianceExample} and \eqref{CallanS2}. Equation \eqref{CallanS2} is a direct generalization of the RG Eq. \eqref{CallanS} to the case of two fields. But Eq. \eqref{quantumscaleinvarianceExample} is the anomaly-free analog of (\ref{regularscaling}-\ref{dimensionalanalysis}): this time there is no RG-related operator on the RHS of the scale transformation and no $\mu_0$ around. One could still relate $\frac{\partial}{\partial t}$ and $\sigmavev\frac{\partial}{\partial \sigmavev}$ in a broken phase of the model, and then to combine the two equations, but we do not present it here, see: \cite{Tamarit:2013vda}. Within the framework of a quantum scale invariant model constructed with the help of scale invariant dimensional regularization, we interpret the C-S equation \eqref{CallanS2} as an expression of the assumed functional invariance of Green functions with respect to choice of the function $\mu(\Phi)$. In proposing the potential \eqref{potential}, we have instrumentally used the simple function
\begin{gather}
\mu(\Phi_1; t) = e^t \Phi_1^{\frac{1}{1-\epsilon}}\Big|_{\Phi_1=\sigmavev + \sigma}
\end{gather}
that depends only on one field and one parameter $t$, so that
\begin{equation}
\mu\frac{\delta}{\delta \mu} \sim \frac{\mathrm{d}}{\mathrm{d}t} = \frac{\partial}{\partial t} + \ldots
\end{equation}
One may imagine taking a more complicated \textit{ansatz} as a point of departure, e.g. $\mu^2 = e^t\left(\Phi_1^2 + e^u \Phi_2^2\right)^{\frac{1}{1-\epsilon}}$, that would give rise to the notion of $u$-dependent running couplings and, we hypothesize, produce additionally an analog of Eq. \eqref{CallanS2} with $\partial/\partial u$, still without introducing the trace anomaly. Thanks to the fact that $\mu$ is an arbitrary function up to the constraints that it has dimension one and depends only on spurious dimensionful parameters, Eq. \eqref{quantumscaleinvarianceExample} can be satisfied despite the presence of nontrivial RG functions in \eqref{CallanS2}.

\section{Special conformal transformations}
\label{special-conformal-section}
In the previous sections of this chapter we have been considering the Poincare algebra $\left\{P^\mu,M^{\alpha\beta}\right\}$,
\begin{gather}
\begin{split}
[P_\mu, P_\nu] =& \;0
\\
[M_{\alpha\beta},P_\mu] =& - i g_{\alpha\mu}P_\beta + ig_{\beta\mu}P_{\alpha}
\\
[M_{\mu\nu},M_{\alpha\beta}] =& -(\mu\alpha,\nu\beta)-(\nu\beta,\mu\alpha) 
\\
&+ (\mu\beta,\nu\alpha) + (\nu\alpha,\mu\beta) \;,
\\
&\text{where}\quad
(\mu\alpha,\nu\beta)\equiv i\eta_{\mu\alpha}M_{\nu\beta}
\end{split}
\end{gather}
extended by the dilatation generator
\begin{gather}
[P_\mu,D] = i P_\mu
\;,\quad
[M_{\mu\nu},D] = 0\,.
\end{gather}
Notice that $\{P^\mu,M^{\alpha\beta},D\}$ closes in the sense that all the above commutators also belong to the algebra.

The particular representation of this algebra that we were interested in was given by the following set of differential operators that act on a function $G^N(x_1,\ldots,x_n)$ interpreted as the $N$-point Green function constructed from scalar quantum fields of dimension $d_\phi$,
\begin{gather}
\label{DilRepr1}
P_\mu = 
i \sum_{i=1}^{N} \frac{\partial}{\partial x_i^\mu}
\;,\quad
M^{\mu\nu} = 
i \sum_{i=1}^N \left(
x_i^\mu \frac{\partial}{\partial {x_i}_\nu} - x_i^\nu \frac{\partial}{\partial {x_i}_\mu}
\right)
\\
\label{DilRepr2}
D = 
i\left(
Nd_\phi + \sum_{i=1}^Nx_i^\mu \frac{\partial}{\partial {x_i}_\mu} 
\right)
\end{gather}
The meaning behind the statement that $G^N$ is Poincare invariant is simply that $P_\mu G^N = 0 = M^{\mu\nu}G^N$. We have also defined the property of quantum scale invariance (in flat space-time) as $DG^N=0$ and we have observed that a generic $\lambda\phi^4$ model does not exhibit this property but a model regularized in the scale-invariant way, \eqref{quantumscaleinvarianceExample}, does.

Further, the $\{P^\mu,M^{\alpha\beta},D\}$ algebra can be extended by four so called special conformal generators $K^\nu$ with the following commutators,
\begin{gather}
\begin{split}
&[K^\nu,D] = -i K^\nu
\;,\quad
[P^\mu,K^\nu] = 2i(\eta^{\mu\nu} D - M^{\mu\nu})
\\
&[K^\mu,M^{\alpha\beta}] = i\eta^{\mu\alpha}K^\beta - i \eta^{\mu\beta}K^\alpha
\;,\quad
[K^\mu,K^\nu] = 0
\end{split}
\end{gather} 
In parallel, the representation (\ref{DilRepr1}-\ref{DilRepr2}) is to be extended by the operator
\begin{equation}
K^\nu = i \left[
\sum_{i=1}^N(2x_i^\nu x_i^\lambda - \eta^{\nu\lambda} x_i^2) \frac{\partial}{\partial x_i^\lambda} + 2 \sum_{i=1}^N d_\phi x_i^\nu
\right]
\label{DilRepr3}
\end{equation}
(The last sum in $K^\nu$ vanishes since the factor $d_\phi$ is constant for all fields in the Green function and $\sum_i x_i = 0$, but it need not to be so in a more general model.)

It is a matter of elementary calculation to check that the differential operators (\ref{DilRepr1}-\ref{DilRepr2},\ref{DilRepr3}) realize the algebra defined by the commutators. Also, again, the algebra $\{P^\mu,M^{\alpha\beta},D,K^\nu\}$ is closed.

At this point it is natural to ask if quantum special conformal invariance of $G^N$, i.e.
\begin{equation}
K^\nu G^N =  0
\end{equation}
strictly accompanies the $DG^N=0$ property. The answer is yes and we show this for the models \eqref{lambdaphitothefourth} and \eqref{thelagrangian2} in a way that is completely analogous to our discussion of dilatations.

Take the following composite operator (that is also a current) $C_\mu^{\;\nu}$ and define new $[\mathcal{O}^S]^\nu$ and $[\mathcal{O}^B]^\nu$ operators
\begin{align}
C_\mu^{\;\nu}
&= [T_{\mu\lambda}](2x^\lambda x^\nu - \eta^{\lambda\nu} x^2)
\\
\begin{split}
\partial^\mu C_\mu^{\;\nu}
&=
(2x^\lambda x^\nu - \eta^{\lambda\nu} x^2)\partial^\mu[T_{\mu\lambda}] + 2x^\nu[T^\mu_{\;\;\mu}]
\\
&\equiv [\mathcal{O}^S]^\nu + [\mathcal{O}^B]^\nu
\end{split}
\end{align}
Using the expressions for $\partial_\mu T^{\mu}_{\;\;\nu}$ and $T^\mu_{\;\;\mu}$, (\ref{TmumuReg}-\ref{dmuTmunuReg}), along with Eqs. (\ref{Edefinition}-\ref{Emudefinition}), we have
\begin{gather}
[\mathcal{O}^S]^\nu \equiv (2x^\lambda x^\nu - \eta^{\lambda\nu} x^2)[E_\lambda] + 2x^\nu d_\phi[E] \\
K^\nu \ast = -\int\!\mathrm{d}^d x\,[\mathcal{O}^S]^\nu\ast = \int\!\mathrm{d}^d x\,[\mathcal{O}^B]^\nu\ast
\end{gather}
Now, for the generic $\lambda\phi^4$, \eqref{lambdaphitothefourth}, we again make use of \eqref{phi4renormalizationEND} obtaining
\begin{gather}
[\mathcal{O}^B]^\nu = 2x^\nu\,2\varepsilon \frac{\lambda_0}{4} \phi_0^4 = 2x^\nu\left(
\gamma_\phi[E]
-\frac{\beta}{4} \mu^{2\varepsilon}[\phi^4]
\right)\
\\
K^\nu\ast \,=\,
- 2i\sum_{i=1}^Nx_i^\nu \gamma_\phi\ast
\,-\,
2\int\!\mathrm{d}^dx\,x^\nu\frac{\beta}{4}\mu^{2\varepsilon}[\phi^4](x)\ast
\end{gather}
Notice that the last contribution to $K^\nu$ above cannot be recast as a simple operator akin of $\beta \frac{\partial}{\partial \lambda}$.

On the other hand, for the quantum scale invariant model \eqref{thelagrangian2} we simply retrieve again, from (\ref{TmumuScaleInvariant}-\ref{dmuTmunu}), that
\begin{equation}
[\mathcal{O}^B]^\nu = 0\;,\quad K^\nu G^N = 0
\end{equation} 
This illustrates how the property of quantum special conformal invariance is realized in tandem with quantum scale invariance as a consequence of vanishing of the trace anomaly, compare equations \eqref{TmumuReg} and \eqref{TmumuScaleInvariant}.

We have paid more attention to the quantum dilatation invariance in this article because it is enough to guarantee the instrumental property of the effective potential being homogeneous, \eqref{Vhomogeneous}.

\section{Summary}
We have introduced the notion of a constraint applied to the space of nonrenormalizable models that allows for the renormalization scale and suppression scale of higher dimensional operators to be interpreted as background field values. We have argued that it might not be possible for renormalizable models to fulfill this requirement. We have also given a simple example of a construction, where the constraint is implemented by specifying nonrenormalizable interactions and using dimensional regularization in a specific way. We have examined in detail how nonperturbative scale symmetry of Green functions is maintained in the presented model thanks to the method of regularization, despite the Callan-Symanzik equation being satisfied as well. The main implication of our analysis is that there exists a broad class of extensions of any field theory where a scalar field has a nonzero vacuum expectation value, to an exactly scale-symmetric theory where the same field's vev lies on an exactly flat direction in the effective potential, and where there are no dimensionful parameters other than vacuum expectation values of fields.

\begin{acknowledgments}
We thank Dumitru Ghilencea for useful comments.  Z.L. was supported by the Polish NCN Grant No. DEC-2012/04/A/ST2/00099. P.O. was supported by the Polish NCN Grant No. DEC-2016/21/N/ST2/03312.
\end{acknowledgments}

\appendix*
\section{Energy momentum tensor}
\label{appendix}


In order for the paper to be self-contained, we aim here to concisely summarize the two main equations that the intuitive meaning behind the notion of the energy-momentum (EM) tensor derives from: the Noether theorem and the Einstein equation.

Consider what is called a field transformation, a function in the space of fields $\Phi \to \Phi'$, that is a combination of a linear function (operator) $T$ and a change in the space-time reference frame $x \to x'$,
\begin{gather}
\begin{split}
x &\,\longrightarrow\, x'(x) = x + \delta x \\
\Phi(x) &\,\longrightarrow\, \Phi'(x'(x)) = T\Phi(x)
\end{split}\label{tranformation}
\\
\Phi'(x') = T\Phi(x(x'))
\label{Phiprime}
\end{gather}
and the difference between $\Phi'(x')$ and $\Phi(x)$ is correspondingly split into two parts,
\begin{align}
\delta &\equiv \delta_0 \,+\, \delta x^\mu\partial_\mu
\\
\begin{split}
\delta \phi(x)
&\equiv(T-\mathds{1})\Phi(x)
\\
&= \underbrace{\Phi'(x) - \Phi(x)}_{\delta_0 \Phi(x)} \,+\, \delta x^\mu\, \partial_\mu \Phi'(x) \,+\, \mathcal{O}(\delta x^2)
\label{delta0}
\end{split}
\end{align} 
We are interested in symmetry transformations, that is those where the two functions in \eqref{tranformation} are representations of the same group. Given an action functional $S[\phi]$, one obtains equations of motions from the condition that $S$ be stationary at physical trajectories in the field space. When the action is invariant with respect to the field change produced by the transformation \eqref{Phiprime}, that is when
\begin{equation}
S[\Phi']\,=\, S[\Phi] \;,
\end{equation}
then each solution to the equations of motion belongs to an orbit of the symmetry group. In other words, \eqref{Phiprime} transforms one solution into another.

Armed with the above notation we can now state the Noether theorem. For simplicity we will focus exclusively on scalar fields. Denote by $\theta^a$ the infinitesimal transformation parameter with an unspecified  set of indeces,
\begin{equation}
\delta x^\mu \,=\, \frac{\delta x^\mu}{\delta\theta^a}\delta \theta^a \;\quad \delta \phi = \frac{\delta\phi}{\delta\theta^a} \delta \theta^a
\label{theta1}
\end{equation}
and assume that the action is given as space-time integral of a local Lagrangian $\mathcal{L}$ that is a function of fields and their first derivatives $\mathcal{L} = \mathcal{L}(\phi, \partial \phi)$. The following expression is the Noether current
\begin{gather}
J^\mu_a \,\equiv\, {T_C}^\mu_{\;\;\nu} \frac{\delta x^\nu}{\delta \theta^a} \,-\, \Pi^\mu \frac{\delta \phi}{\delta \theta^a}
\;,\text{where}\\
\Pi^\mu \,=\,\frac{\partial \mathcal{L}}{\partial (\partial_\mu \phi)}\;,\quad
{T_C}^{\mu\nu} = \Pi^\mu \partial_\nu \phi - g^\mu_{\;\;\nu} \mathcal{L}
\end{gather}
and $T_C$ is the canonical energy-momentum (EM) tensor. When the system is symmetric, the Noether current is conserved,
\begin{equation}
\partial_\mu J^\mu_a \,=\, 0\, .
\label{Noetherconservation}
\end{equation}
A nonzero right-hand side above serves as an indication why and to what degree is the symmetry violated.

For the transformation that consists only of space-time translation, $\delta x^\mu \,=\, \epsilon^\mu$, $\delta \phi \,=\, 0$, the EM tensor is the whole Noether current, and it is conserved as the consequence of translation invariance,
\begin{equation}
J^\mu_{\nu} = {T_C}^\mu_{\;\;\rho} \delta^\rho_{\;\;\nu}
\;,\quad \partial_\mu {T_C}^{\mu\nu} \,=\,0\,.
\label{EMtensorasNoether}
\end{equation}
The canonical tensor also features in the conserved charges (generators) $\int\!\mathrm{d}^3\vec{x} \,J_a^0(x)$ for the Poincare group
\begin{gather}
P^\mu \,\equiv\, \int\!\mathrm{d}^3\vec{x} \,{T_C}^{0\mu}
\\
M_{\alpha\beta} \,=\,\int\!\mathrm{d}^3\vec{x}
\left(
{T_C}^0_{\;\;\alpha}x_\beta - 
{T_C}^0_{\;\;\beta}x_\alpha
\right)
\end{gather}

Now we introduce the modified EM tensor $T$, along with a free parameter $\eta$ (and $\Delta \eta$),
\begin{gather}
T^{\mu\nu} \,\equiv\, {T_C}^{\mu\nu} - \eta\, (\partial_\mu\partial_\nu - g_{\mu\nu}\partial^2 ) \phi^2
\label{EMtensormod}
\\
\eta\,\equiv\,\frac{1}{2(d-1)} \left(\Delta \eta + d_\phi\right)\, \in\,\mathbb{R}
\end{gather}
$T^{\mu\nu}$ is just as good EM tensor as $T_C$ in the sense that $\partial_\mu T^{\mu\nu} = \partial_\mu {T_C}^{\mu\nu} = 0$, and that $T$ can be swapped for $T_C$ in the generators $P^\mu$ and $M_{\alpha\beta}$ (up to boundary terms).

Finally we turn to the (classical) scale symmetry transformation, which is defined (in $d$ dimensions) by
\begin{align}
&x' = e^s x\;,\quad 
\delta x^\mu = s x^\mu 
\label{scaletransformation1}
\\
\begin{split}
&\phi'(x') = e^{-sd_\phi} \phi(x) \;,\quad
\delta \phi = -sd_\phi \phi
\\
& \quad d_\phi=(d-2)/2
\end{split}
\\
&\phi'(x) = e^{-sd_\phi} \phi(e^{-s}x)
\label{scaletransformation3}
\end{align}
We also specify our discussion only to Lagrangians of the form $\mathcal{L} = \frac{1}{2}\left(\partial \phi\right)^2 - V(\phi)$. Thus, the related Noether current $J_s$ is given by
\begin{gather}
{J_s}^\mu \,=\, x_\nu\,{T_C}^{\mu\nu} \,+\, d_\phi \phi \partial^\mu \phi \\
{T_C}^{\mu\nu} = \partial^\mu\phi \partial^\nu\phi - g^{\mu\nu} \mathcal{L}
\end{gather}
By noticing that 
\begin{gather}
\begin{split}
x_\nu({T_C}^{\mu\nu} - T^{\mu\nu}) 
= \eta \, x_\nu(\partial^\mu\partial^\nu - g^{\mu\nu} \partial^2) \phi^2
\\
= \eta \,\partial_\nu(x^\nu \partial^\mu - x^\mu\partial^\nu)\phi^2 - 2\eta \,(d-1) \phi\partial_\nu \phi
\end{split}
\end{gather}
we can write
\begin{gather}
D^\mu \equiv {J_s}^\mu + \eta \partial_\nu(x^\mu \partial^\nu - x^\nu\partial^\mu)\phi^2 \\
D^\mu= x_\nu T^{\mu\nu} - \Delta \eta \,\phi\partial^\mu \phi
\end{gather}
$D^\mu$ is the dilatation current. The integrals
$
D\equiv \int\!\mathrm{d}^3\vec{x}\,D^0 = 
\int\!\mathrm{d}^3\vec{x}\,{J_s}^0
$
that give the dilatation generator are equal up to a boundary term. In a scale symmetric theory
\begin{equation}
\partial_\mu D^\mu = \partial_\mu {J_s}^\mu = 0
\end{equation}
and in any case, using \eqref{EMtensorasNoether},
\begin{equation}
\partial_\mu D^\mu = T^\mu_{\;\;\mu} - \frac{\Delta \eta}{2}\partial^2 \phi^2\;.
\label{dD}
\end{equation}
On the other hand, when we substitute our posited Lagrangian with the canonical kinetic term we get
\begin{equation}
\partial_\mu D^\mu = d\, V + d_\phi\, \phi\partial^2 \phi  = \left(d - d_\phi\, \phi \frac{\partial}{\partial \phi} \right) V\;,
\label{dilatationfourdivergence}
\end{equation}
where in the last step we have substituted the equation of motion $\partial^2 \phi + V'(\phi) = 0$. As indicated earlier, the four-divergence of dilatation current is given by scale symmetry violating terms. The latter in turn are given in the simplest case by inhomogeneity of the potential as a function of the scalar field. Thus clearly a mass term $m^2 \phi^2$ or nonrenormalizable terms like $\frac{\phi^6}{M^2}$ would contribute to the right-hand side of \eqref{dilatationfourdivergence} in $d=4$ dimensions. The main discussion in the article adds nuance to this statement.

Now we turn to the theory of relativity to shed light on the $\Delta \eta$ parameter in \eqref{dD}. Consideration anywhere in the article but in this paragraph assume that the metric is flat, $g_{\mu\nu} = \eta_{\mu\nu}$. Now consider gravity. Couple the scalar field to the metric in standard way,
\begin{equation}
S = \int\!\!\sqrt{-g}\, \left[ \frac{1}{2} g^{\mu\nu} \partial_\mu \phi\partial_\nu \phi - V(\phi) - \frac{1}{2}\left(\eta\phi^2 + M_P^2 \right) R\right]
\end{equation}
Demanding that $S$ be invariant with respect to metric variations, one obtains Einstein equations ($\nabla$ is the covariant derivative),
\begin{gather}
\begin{split}
0 &= \frac{1}{\sqrt{-g}} \frac{\delta S\;}{\delta g^{\mu\nu}}
\\
&=
-\frac{1}{2}\left(R_{\mu\nu} - \frac{1}{2}g_{\mu\nu}R\right)\left(\eta\phi^2 + M_P^2\right) - \frac{1}{2} T_{\mu\nu} 
\label{Einsteineq}
\end{split}
\\
\begin{split}
T_{\mu\nu} \equiv 
&\;\partial_\mu \phi \partial_\nu \phi
- g_{\mu\nu} \left(
\frac{1}{2}(\partial \phi)^2 - V
\right)
\\ 
&- \eta(\nabla_\mu\nabla_\nu - g_{\mu\nu} \nabla^2)\phi^2
\end{split}
\end{gather}
This $T^\mu_{\;\;\nu}$ coincides in the flat limit with the energy-momentum tensor introduced in \eqref{EMtensormod}. But then, seeing as $R=0$, the choice of $\eta$ is ambiguous.

Examine the trace of \eqref{Einsteineq},
\begin{align}
T^\mu_{\;\;\mu} 
&= 
\left(1- \frac{d}{2}\right)(\partial \phi)^2 + d V + 2\eta(d-1)\left[(\partial\phi)^2 + \phi \nabla^2 \phi \right]
\nonumber
\\
&=
\left(1- \frac{d}{2}\right) R (\eta \phi^2 + M_P^2)
\end{align}
Plug in our definition of $\eta$ in terms of $\Delta \eta$, and the equation of motion for $\phi$, $
\phi\nabla^2 \phi + \phi\frac{\partial}{\partial \phi} V + \eta \phi^2 R = 0
$
and arrive at
\begin{equation}
\left(1- \frac{d}{2}\right)R M_P^2 = \left(d - d_\phi \phi \frac{\partial}{\partial \phi} \right) V + \frac{\Delta \eta}{2} \nabla^2 \phi^2
\end{equation}
A hypothetical choice of $M_P^2=0$ and of a potential $V(\phi)$ that is a homogeneous function, $(d -d_\phi \phi \,\partial/\partial \phi)V = 0$, is equivalent to forbidding any dimensionful parameters in the Lagrangian, resulting in the presence of classical scale symmetry. In such case we get
\begin{equation}
 0 = \Delta \eta\, \nabla^2 \phi^2 = \Delta\eta\,\nabla^\mu(\partial_\mu\phi^2)
\end{equation}
and this puts us in front of two distinct possibilities, either $\Delta \eta \neq 0$ or $\Delta \eta = 0$. In the first case the current $\partial_\mu\phi^2$ is conserved in curved space-time. (
This can be consequential, see e.g. the exhaustive work in \cite{GarciaBellido:2011de,Bezrukov:2012hx,Karananas:2016kyt} and later in \cite{Ferreira:2016vsc,Ferreira:2016wem,Ferreira:2016kxi,Ferreira:2018itt,Ferreira:2018qss}.
)
But it is not uncommon for Authors, e.g. \cite{Pokorski:1987ed,Ramond:1981pw}, to choose the second possibility 
\begin{equation}
\Delta\eta=0 
\;,\quad
 \eta = \frac{d-2}{8(d-1)}
\label{etacondition}
\end{equation}
This leaves one with a pleasantly consistent double interpretation of the trace $T^\mu_{\;\;\mu}$. It is now both the source of space-time curvature and the measure of scale symmetry violation as dictated by \eqref{dD}.

A related significant characteristic of choosing the point \eqref{etacondition} (and forbidding dimensionful parameters) in curved space-time is that it results in the action
\begin{equation}
S = \int\!\!\sqrt{-g}\left[
\frac{1}{2} g^{\mu\nu} \partial_\mu \phi \partial_\nu \phi - V(\phi) - \frac{d_\phi}{4(d-1)}\phi^2 R
\right]
\end{equation}
that is invariant not only with global but also local scale symmetry, $s=s(x)$, implemented with a transformation of the metric rather than the coordinates $x$, compare (\ref{scaletransformation1}-\ref{scaletransformation3}),
\begin{align}
g_{\mu\nu}(x) \,&\longrightarrow\, g'_{\mu\nu}(x) = e^{2s(x)}g_{\mu\nu}(x)
\\
\phi(x) \,&\longrightarrow\, \phi'(x) = e^{-s(x)d_\phi}(x)
\end{align}
also known as Weyl symmetry.

In summary, $\eta$ is interpreted as the coupling between curvature $R$ and a scalar field, that imprints itself on the EM tensor and remains present in the flat space-time. We do not go further into discussion of coordinate invariance or Weyl symmetry, but we will stay focused on global scale transformation in flat space-time and its possible violation by quantum effects. Generically the $\eta$ parameter is needed for regularization and subject to renormalization. From this point of view the choice in \eqref{etacondition} is an arbitrary but convenient renormalization condition. The convenience stems from the fact that it is stable with respect to including loop corrections \cite{Brown:1979pq}. The resulting final expression for the (classical) dilatation current is simply
\begin{equation}
D^\mu \,=\, x_\nu T^{\mu\nu} \,.
\end{equation}

When the field becomes a quantum operator, so does the transformation in \eqref{Phiprime}. It is (most often) implemented with a unitary operator $U$ (and hermitian 
$G$)
\begin{gather}
\Phi \,\longrightarrow\, \Phi' = U \Phi U^{-1} = e^{i\theta G}\Phi e^{-i\theta G} \\
\delta_0 \Phi \,=\, i\theta [G,\Phi]\;
\end{gather}
We equate the above $\delta_0$ with the one defined previously in \eqref{delta0}. Take the infinitesimal translation and dilatation, and the corresponding generators $P_\mu$ and $D$, acting on a scalar field $\phi$,
\begin{align}
\delta_0^P \phi &= - \epsilon^\mu \partial_\mu \phi
\stackrel{!}{=}
i\,\epsilon^\mu\left[P_\mu,\Phi\right] 
\\
\delta_0^D\Phi &= -s (d_\phi + x^\mu \partial_\mu )\Phi \stackrel{!}{=} i\,s\left[D,\Phi\right]
\end{align}
Further, we want the resulting algebras to be consistent, meaning that the commutator of $D$ and $P$ can be obtained as
\begin{align}
\begin{split}
&s\epsilon^\mu\left[ iD,iP_\mu \right]
\stackrel{!}{=}
\left[\delta_0^D, \delta_0^P \right]
\\
&
\;= \left[
-s\left(d_\phi + x^\mu\partial_\mu\right), -\epsilon^\nu\partial_\nu\right]
=-s\epsilon^\nu \partial_\nu
\stackrel{!}{=} is\epsilon^\nu P_\nu 
\end{split}
\\
&\Rightarrow i[D,P_\mu] = P_\mu
\label{DPkomutator}
\end{align}
Casting the classical physics aside, \eqref{DPkomutator} is the defining property of the Poincare algebra extended by the dilatation generator. The other new commutators are zero, $[D,M_{\alpha\beta}] = 0$. The maximal bosonic extension of the Poincare algebra also includes generators of the special conformal transformations, $K^\mu$. The main body of the text focuses on dilatations but the discussion can be extended to encompass $K^\mu$ and special conformal transformations of Green functions, see Sec. \ref{special-conformal-section}.

\bibliography{Vanishing_trace_anomaly_bibliography}

\end{document}